\documentclass[reqno,12pt]{amsart}

\usepackage{amsmath,amssymb,amsthm,amsfonts,amsxtra}
\usepackage{fullpage}
\usepackage{graphicx}
\usepackage{caption}
\usepackage{subcaption}
\usepackage{float}

\begin{document}

\newtheorem{thm}{Theorem}[section]
\newtheorem{cor}{Corollary}[section]
\newtheorem{lem}{Lemma}[section]
\newtheorem{prop}{Proposition}[section]

\def\const{\text{const.}}
\def\Rnum{{\mathbb R}}
\def\sgn{{\rm sgn}}
\def\Dop{{\mathcal{D}}}
\def\nvec{{\mathbf n}}
\def\kvec{{\mathbf k}}
\def\sech{{\rm sech}}

\def\sig{\sigma^2}
\def\sqrtsig{\sigma}

\tolerance=10000
\allowdisplaybreaks[3]

\title{Conservation laws and line soliton solutions of \\a family of modified KP equations}

\author{
Stephen C. Anco$^1$
\lowercase{\scshape{and}}
M.L. Gandarias${}^2$
\lowercase{\scshape{and}}
Elena Recio$^{2}$ 
\\\\\lowercase{\scshape{
${}^1$Department of Mathematics and Statistics\\
Brock University\\
St. Catharines, ON L$\scriptstyle{2}$S$\scriptstyle{3}$A$\scriptstyle{1}$, Canada}} \\\\
\lowercase{\scshape{
${}^2$Department of Mathematics\\
Faculty of Sciences, University of C\'adiz\\
Puerto Real, C\'adiz, Spain, $\textstyle{11510}$}}\\
}


\begin{abstract}
A family of modified Kadomtsev-Petviashvili equations (mKP) in 2+1 dimensions is studied.
This family includes the integrable mKP equation 
when the coefficients of the nonlinear terms and the transverse dispersion term 
satisfy an algebraic condition. 
The explicit line soliton solution and all conservation laws of low order are derived for all equations in the family  
and compared to their counterparts in the integrable case. 
\end{abstract}

\maketitle

\section{Introduction}

In 2+1 dimensions, an integrable generalization of the modified Korteweg-de Vries (mKdV) equation 
is the modified Kadomtsev-Petviashvili (mKP) equation \cite{KonDub1984}
\begin{equation}\label{mKP}
(u_t -\alpha u^2 u_x \pm \sqrt{2\alpha\gamma} u_x\partial_x^{-1}u_y +\beta u_{xxx})_x +\gamma u_{yy}=0
\end{equation}
where $\alpha,\beta,\gamma$ are non-zero constants.
This equation arises in several physical applications \cite{CheLiu,DasSar,VeeDan} 
pertaining to dispersive nonlinear wave phenomena. 
Unlike the better known Kadomtsev-Petviashvili (KP) equation \cite{KadPet},
the mKP equation contains a nonlocal term and has no obvious Lagrangian structure. 
Its line soliton solutions and some conservation laws can be found in 
Refs.~\cite{GesHolSaaSim,KonDub1992,NazAliNae,ZhaXuJiaZho}. 

In the present paper,
we consider a family of mKP equations 
\begin{equation}\label{mKP-fam}
(u_t -\alpha u^2 u_x +\kappa u_x\partial_x^{-1}u_y +\beta u_{xxx})_x +\gamma u_{yy}=0
\end{equation}
with arbitrary non-zero constant coefficients $\alpha,\beta,\gamma,\kappa$. 
The integrable mKP equation is given by the case 
\begin{equation}\label{mKP-constraint}
\kappa^2=2\alpha\gamma >0 . 
\end{equation}
This family \eqref{mKP-fam} can be expected to have wider applications 
in physical situations where the integrability constraint \eqref{mKP-constraint} does not hold. 

The main goals will be 
to determine the line soliton solutions and the low-order conservation laws of the mKP family \eqref{mKP-fam}
and to compare the results to the integrable mKP case. 
First, in section~\ref{potential},
the mKP family \eqref{mKP-fam} is formulated as a local PDE 
by use of the potential $w$, with $u=w_x$. 

Next, in section~\ref{conslaws}, 
all low-order conservation laws of the mKP family in potential form are derived. 
The admitted conservation laws are found to consist of 
two topological charges, for arbitrary $\kappa$, 
plus two additional topological charges 
in the integrable case \eqref{mKP-constraint} of the mKP equation. 
Computational aspects are summarized in an appendix.
Unlike the KP equation, the mKP family in potential form does not admit any
non-trivial dynamical conserved quantities. 

In section~\ref{solns}, 
all line solitons $u=U(x+\mu y-\nu t)$ are derived, 
where the parameters $\mu$ and $\nu$ determine the direction and the speed of the line soliton. 
The basic kinematical properties of these solutions are discussed
and compared to the mKP line solitons.
In particular, there is a significant qualitative difference between
the (extended) mKP case where $\alpha\gamma >0$
and the opposite case where $\alpha\gamma<0$. 

Finally, a few concluding remarks are made in section~\ref{remarks}.

\section{Potential form}\label{potential}

The mKP family \eqref{mKP-fam} is equivalent to a local PDE system 
\begin{equation}\label{mKP-sys}
u_t -\alpha u^2 u_x +\kappa u_x v +\beta u_{xxx} +\gamma v_y=0,
\quad
v_x = u_y . 
\end{equation}
This system can be expressed as a single PDE 
by the introduction of a potential $w$ given by 
\begin{equation}\label{pot}
u=w_x,
\quad
v=w_y, 
\end{equation}
which yields 
\begin{equation}\label{mKP-pot}
0 = w_{tx} -\alpha w_x^2 w_{xx} +\kappa w_{xx} w_y +\beta w_{xxxx} +\gamma w_{yy} .
\end{equation}
By applying a general scaling transformation
$t\to \lambda_1 t$, $x\to \lambda_2 x$, $y\to \lambda_3 y$, $w\to \lambda_4 w$,
where $\lambda_1,\lambda_2,\lambda_3,\lambda_4\neq0$,
we can fix three of the four coefficients $\alpha,\beta,\gamma,\kappa$:
$|\alpha| =\beta =|\gamma| =1$; 
also we can fix the sign $\kappa>0$, 
without loss of generality.

Hence, we will consider the mKP family in the scaled potential form 
\begin{equation}\label{mKP-pot-scaled}
0 = w_{tx} + (\sigma_1 w_x^2 +\kappa w_y) w_{xx} + w_{xxxx} +\sigma_2 w_{yy},
\quad
\sigma_1,\sigma_2=\pm 1,
\quad
\kappa>0, 
\end{equation}
which is a one-parameter family 
where $\kappa$ (rescaled) is an arbitrary positive constant.
We will refer to $\sigma_1=1$ as the focussing case, and $\sigma_1=-1$ as the defocussing case;
this distinction will be significant when line soliton solutions are considered. 

The corresponding scaled mKP family has the form 
\begin{equation}\label{mKP-fam-scaled}
(u_t +\sigma_1 u^2 u_x +\kappa u_x\partial_x^{-1}u_y + u_{xxx})_x +\sigma_2 u_{yy}=0,
\quad
\sigma_1,\sigma_2=\pm 1,
\quad
\kappa>0
\end{equation}
in which the scaled mKP equation is the case  
\begin{equation}\label{mKP-case}
\kappa^2=2,
\quad
\sigma_1\sigma_2 =-1,
\end{equation}
namely, 
\begin{equation}\label{mKP-scaled}
(u_t +\sigma_1 u^2 u_x +\sqrt{2} u_x\partial_x^{-1}u_y + u_{xxx})_x -\sigma_1 u_{yy}=0,
\quad
\sigma_1=\pm 1 .
\end{equation}

\section{Conservation laws}\label{conslaws}

Conservation laws are of basic importance for nonlinear evolution equations 
because they provide physical, conserved quantities as well as conserved norms.
A general treatment of how to find conservation laws is given in 
Refs.~\cite{Olv-book,AncBlu2002b,BCA-book,Anc-review}.

For the mKP family in potential form \eqref{mKP-pot-scaled}, 
a local conservation law is a continuity equation
\begin{equation}\label{conslaw}
D_t T+D_x X+D_y Y=0
\end{equation}
holding for all solutions $w(x,y,t)$ of equation \eqref{mKP-pot-scaled},
where $T$ is the conserved density, and $(X,Y)$ is the spatial flux,
which are functions of $t$, $x$, $y$, $w$, and derivatives of $w$. 
When solutions $w(x,y,t)$ are considered
in a given spatial domain $\Omega\subseteq\Rnum^2$, 
every local conservation law yields a corresponding conserved integral 
\begin{equation}\label{conservedintegral}
\mathcal{C}[w]= \int_{\Omega} T\,dx\,dy
\end{equation}
satisfying the global balance equation
\begin{equation}\label{globalconslaw}
\frac{d}{dt}\mathcal{C}[w]= -\int_{\partial\Omega} (X,Y)\cdot\hat\nvec\,ds
\end{equation}
where $\hat\nvec$ is the unit outward normal vector of the domain boundary curve $\partial\Omega$, 
and where $ds$ is the arclength on this curve with clockwise orientation. 
This global equation \eqref{globalconslaw} has the physical meaning that
the rate of change of the quantity \eqref{conservedintegral} on the spatial domain 
is balanced by the net outward flux through the boundary of the domain. 

A conservation law is locally trivial \cite{Olv-book,BCA-book,Anc-review} 
if, for all solutions $w(x,y,t)$ in $\Omega$,
the conserved density $T$ reduces to a spatial divergence $D_x \Psi^x + D_y \Psi^y$ 
and the spatial flux $(X,Y)$ reduces to a time derivative $-D_t(\Psi^x,\Psi^y)$ 
modulo a spatial curl $(D_y\Theta,-D_x\Theta)$, 
since then the global balance equation \eqref{globalconslaw} becomes an identity. 
Likewise, two conservation laws are locally equivalent \cite{Olv-book,BCA-book,Anc-review} 
if they differ by a locally trivial conservation law, for all solutions $w(x,y,t)$ in $\Omega$. 
We will be interested only in locally non-trivial conservation laws. 

Any non-trivial conservation law \eqref{conslaw} 
can be expressed in an equivalent characteristic form \cite{Olv-book,BCA-book,Anc-review}
which is given by a divergence identity holding off of the space of solutions $w(x,y,t)$. 
For the mKP family in potential form \eqref{mKP-pot-scaled}, 
conservation laws have the characteristic form
\begin{equation}\label{mKP-chareqn}
D_t\tilde T+D_x\tilde X+D_y\tilde Y=
( w_{tx} + (\sigma_1 w_x^2 +\kappa w_y) w_{xx} +w_{xxxx} +\sigma_2 w_{yy} )Q
\end{equation}
where $\tilde T$, $\tilde X$, $\tilde Y$, and $Q$ 
are functions of $t$, $x$, $y$, $w$, and derivatives of $w$,
and where the conserved density $\tilde T$ and the spatial flux $(\tilde X,\tilde Y)$
reduce to $T$ and $(X,Y)$ when restricted to all solutions $w(x,y,t)$ of equation \eqref{mKP-pot-scaled}. 
This divergence identity is called the characteristic equation for the conservation law,
and the function $Q$ is called the conservation law multiplier. 
Note that, when a conservation law is non-trivial, 
$Q$ will be non-singular when it is evaluated on any solution $w(x,y,t)$. 

From the characteristic form \eqref{mKP-chareqn}, 
all multipliers $Q$ are determined by applying 
the Euler operator \cite{Olv-book,BCA-book,Anc-review} 
$E_w$ with respect to $w$,
where this operator annihilates a function of $t$, $x$, $y$, $w$, and derivatives of $w$
iff the function is given by a total divergence. 
Hence, multipliers $Q$ are the solutions of the determining equation 
\begin{equation}\label{Q-deteqn}
E_w\big(( w_{tx} + (\sigma_1 w_x^2 +\kappa w_y) w_{xx} +w_{xxxx} +\sigma_2 w_{yy} )Q\big) =0
\end{equation}
holding off of solutions of equation \eqref{mKP-pot-scaled}. 
All multipliers $Q$ up to any specified differential order with respect to $w$
can be found by splitting the determining equation \eqref{Q-deteqn} 
with respect to all variables that do not appear in $Q$, 
yielding an overdetermined system to be solved for $Q$. 
A variety of methods \cite{Wol,Anc2003,BCA-book,Anc-review}
can be used to derive the conserved density $\tilde T$ and spatial flux $(\tilde X,\tilde Y)$
arising from any given multiplier $Q$. 

Here we will explicitly find all low-order conservation laws of the mKP family in potential form \eqref{mKP-pot-scaled}
by determining all multipliers of differential order at most three
\begin{equation}\label{low-order-Q}
Q(t,x,y,w,\partial w,\partial^2 w,\partial^3 w) 
\end{equation}
where $\partial^k w$ denotes the set of all partial derivatives of order $k\geq 0$ of $w$.
Some remarks on the computations are provided in the appendix.  

\begin{prop}
All low-order multipliers \eqref{low-order-Q} admitted by 
the potential form of the mKP family \eqref{mKP-pot-scaled} 
with $\kappa\neq0$, $\sigma_1^2=1$, $\sigma_2^2=1$ are given by\\
(i) $\kappa$ arbitrary:
\begin{align}
& \label{mult1}
Q_{1} = f_1(t) ,
\\
& \label{mult2}
Q_{2} = \kappa w_xf_2(t)  + y f'_2(t) ; 
\end{align}
(ii) $\kappa^2=2$, $\sigma_2\sigma_1=-1$:
\begin{align}
& \label{mult3}
Q_{3} = (4 y w_x - 2\kappa\sigma_2 x) f_3(t) +\kappa y^2 f'_3(t) , 
\\
& \label{mult4}
\begin{aligned}
Q_{4} = 
&
- ( \kappa\sigma_1 w_{xxx} 
+\tfrac{1}{3} \kappa w_x^3 
-\sigma_1 w_x w_y
-\tfrac{1}{4} \kappa\sigma_1 w_t )f_{4}(t) 
\\& \qquad
+( \tfrac{1}{4} \kappa\sigma_1 x w_x )f'_{4}(t) 
+( \tfrac{1}{8} \kappa y^2 w_x + \tfrac{1}{4} \sigma_1 xy )f''_{4}(t) 
+ \tfrac{1}{24} y^3 f'''_{4}(t) ; 
\end{aligned}
\end{align}
where $f_1(t),f_2(t),f_3(t),f_4(t)$ are arbitrary functions. 
\end{prop}

These multipliers yield all non-trivial conservation laws of low order,
summarized as follows. 

\begin{thm}\label{conslaws-mkp}
(i) The low-order conservation laws
admitted by the mKP family in potential form \eqref{mKP-pot-scaled} 
for arbitrary $\kappa$
are given by (up to equivalence)
\begin{subequations}\label{mkp-conslaw1}
\begin{flalign}
\quad&\begin{aligned}
T_{1} = &
0
, 
\end{aligned}&&
\\
\quad&\begin{aligned}
X_{1} = & 
f_1(t) \big(
w_{xxx}
+\kappa w_x w_y 
+\tfrac{1}{3}\sigma_1 w_x^3
+w_t
\big)  
, 
\end{aligned}&&
\\
\quad&\begin{aligned}
Y_{1}= & 
f_1(t) \big(
\sigma_2 w_y 
-\tfrac{1}{2}\kappa w_x^2
\big) 
;
\end{aligned}&&
\end{flalign}
\end{subequations}
\begin{subequations}\label{mkp-conslaw2}
\begin{flalign}
\quad&\begin{aligned}
T_{2} = & 
0
, 
\end{aligned}&&
\\
\quad& \begin{aligned}
X_{2} = & 
f_2(t) \big(
(-w_{txxx}-(\sigma_1 w_{x}^2+\kappa w_{y})w_{tx}-\kappa w_{x}w_{ty}-w_{tt})y
+\kappa w_{x}w_{xxx} -\tfrac{1}{2}\kappa w_{xx}^2
\\& \qquad
+\tfrac{1}{4}\kappa w_{x}^4\sigma_1 +\tfrac{1}{2}\kappa^2w_{x}^2w_{y}-\tfrac{1}{2}\kappa \sigma_2 w_{y}^2
\big)
, 
\end{aligned}&&
\\
& \begin{aligned}
Y_{2}= & 
f_2(t) \big(
(\kappa w_{x}w_{tx}-\sigma_2 w_{ty})y-\tfrac{1}{6}\kappa ^2w_{x}^3+\kappa \sigma_2 w_{x}w_{y}+\sigma_2 w_{t}
\big)
;
\end{aligned}
\end{flalign}
\end{subequations}
where $f_1(t)$, $f_2(t)$ are arbitrary functions. 
\\
(ii) Additional low-order conservation laws are admitted 
only when $\kappa^2=2$, $\sigma_2\sigma_1=-1$. 
These conservation laws consist of (up to equivalence):\\
\begin{subequations}\label{mkp-conslaw3}
\begin{flalign}
\quad&\begin{aligned}
T_{3} = & 
0
, 
\end{aligned}&&
\\
\quad&\begin{aligned}
X_{3} = &
f_3(t) \big(
(w_{xxx}+\tfrac{1}{3}\sigma_1 w_{x}^3+\kappa w_{y}w_{x}+w_{t})x
-\tfrac{1}{2}(\sigma_1 w_{txxx}
+(w_{x}^2+\kappa \sigma_1 w_{y})w_{tx} +\sigma_1 w_{tt}
\\& \qquad
+\sigma_1 \kappa w_{x}w_{ty})y^2
+(\sigma_1 w_{x}^2w_{y}+\sigma_1 \kappa w_{x}w_{xxx}-\tfrac{1}{2}\sigma_1 \kappa w_{xx}^2+\tfrac{1}{4}\kappa w_{x}^4+\tfrac{1}{2}\kappa w_{y}^2)y
-w_{xx}
\big)
, 
\end{aligned}&&
\\
\quad&\begin{aligned}
Y_{3}= & 
f_3(t) \big(
(-\sigma_1 w_{y}-\tfrac{1}{2}\kappa w_{x}^2)x
+(\tfrac{1}{2}\sigma_1 \kappa w_{x}w_{tx}+\tfrac{1}{2}w_{ty})y^2
-(w_{t} +\kappa w_{y}w_{x} +\tfrac{1}{3}\sigma_1 w_{x}^3)y
\big) ; 
\end{aligned}&&
\end{flalign}
\end{subequations}
\begin{subequations}\label{mkp-conslaw4}
\begin{flalign}
\quad&\begin{aligned}
T_{4} = &
0
, 
\end{aligned}&&
\\
\quad&\begin{aligned}
X_{4} = & 
f_4(t) \big( 
(-\tfrac{1}{8}\kappa w_{ttt} -\tfrac{1}{8}w_{ttxxx}\kappa
-(\tfrac{1}{4}w_{y}+\tfrac{1}{8}\sigma_1 w_{x}^2\kappa )w_{ttx}
-\tfrac{1}{4}w_{tty}w_{x} -\tfrac{1}{2}w_{tx}w_{ty}
\\& \qquad
-\tfrac{1}{4} \sigma_1 \kappa w_{x}w_{tx}^2 )xy
+( \tfrac{1}{4}w_{x}w_{txxx}-\tfrac{1}{4}w_{txx}w_{xx}
+\tfrac{1}{4}(\kappa w_{y}w_{x}+w_{xxx}+\sigma_1 w_{x}^3)w_{tx}
\\& \qquad
+\tfrac{1}{8}\kappa w_{x}^2w_{ty}+\tfrac{1}{4}\sigma_1 w_{y}w_{ty} )x
+(\tfrac{1}{48}\kappa \sigma_1 w_{tttt}
+\tfrac{1}{48}\kappa\sigma_1 w_{tttxxx}
+(\tfrac{1}{24}\sigma_1 w_{y}+\tfrac{1}{48}\kappa w_{x}^2)w_{tttx}
\\& \qquad
+\tfrac{1}{24}\sigma_1 w_{x}w_{ttty}
+\tfrac{1}{8}(\kappa w_{x}w_{tx}+\sigma_1 w_{ty})w_{ttx}
+\tfrac{1}{8}\sigma_1 w_{tx}w_{tty} +\tfrac{1}{24}\kappa w_{tx}^3 )y^3
\\& \qquad
+( -\tfrac{1}{8}\sigma_1 (w_{ttxxx}w_{x} -w_{ttxx}w_{xx})
-\tfrac{1}{8}(w_{x}^3 +\sigma_1 w_{xxx} +\sigma_1 \kappa w_{y}w_{x})w_{ttx}
-\tfrac{1}{16}\sigma_1\kappa w_{tty}w_{x}^2
\\& \qquad
-\tfrac{1}{8}w_{tty}w_{y}
-\tfrac{1}{4}\sigma_1 w_{txxx}w_{tx}
+\tfrac{1}{8}\sigma_1 w_{txx}^2
-(\tfrac{1}{8}\kappa \sigma_1 w_{y}+\tfrac{3}{8}w_{x}^2)w_{tx}^2
-\tfrac{1}{4}\sigma_1 \kappa w_{tx}w_{ty}w_{x}
\\& \qquad
-\tfrac{1}{8}w_{ty}^2 )y^2
+(\tfrac{1}{8}\kappa w_{ttxx}) y
-\tfrac{1}{4}w_{txx}w_{x}+\tfrac{1}{2}w_{xx}w_{tx}+\tfrac{1}{8}w_{t}^2
+(\tfrac{1}{4}w_{xxx}+\tfrac{1}{12}\sigma_1 w_{x}^3
\\& \qquad
+\tfrac{1}{4}\kappa w_{y}w_{x})w_{t}
+\tfrac{1}{2}w_{xxx}^2
+\tfrac{1}{3}w_{xxx}\sigma_1 w_{x}^3
+(\tfrac{1}{2}\kappa w_{xxx}w_{y}-\tfrac{1}{2}\kappa w_{xx}w_{xy})w_{x}
-\tfrac{1}{2}\sigma_1 w_{xy}^2
\\& \qquad
+\tfrac{1}{4}\kappa w_{xx}^2w_{y} -\sigma_1 w_{xx}w_{yy}
+\tfrac{1}{18}w_{x}^6+\tfrac{5}{24}\kappa \sigma_1 w_{y}w_{x}^4 +\tfrac{1}{2}w_{y}^2w_{x}^2 +\tfrac{1}{12}\kappa \sigma_1w_{y}^3 
\big) 
,
\end{aligned}&&
\\
\quad&\begin{aligned}
Y_{4}= &
f_4(t) \big(
(\tfrac{1}{4}w_{ttx}w_{x}+\tfrac{1}{8}\sigma_1 \kappa w_{tty} +\tfrac{1}{4}w_{tx}^2 )xy
-( (\tfrac{1}{8}\kappa w_{x}^2 +\tfrac{1}{4}\sigma_1 w_{y})w_{tx}
+\tfrac{1}{8}\kappa \sigma_1 w_{tt} +\tfrac{1}{4}\sigma_1 w_{x}w_{ty} )x
\\& \qquad
-(\tfrac{1}{48}\kappa w_{ttty} +\tfrac{1}{24}\sigma_1 w_{tttx}w_{x} +\tfrac{1}{8}\sigma_1 w_{ttx}w_{tx} )y^3
+( \tfrac{1}{16}\kappa w_{ttt}
+(\tfrac{1}{16}\sigma_1 w_{x}^2\kappa +\tfrac{1}{8}w_{y})w_{ttx}
\\& \qquad
+\tfrac{1}{8}w_{tty}w_{x} 
+\tfrac{1}{8}\sigma_1 \kappa w_{tx}^2 w_{x} +\tfrac{1}{4}w_{tx}w_{ty} )y^2
+(-\tfrac{1}{8}\kappa w_{x}^2-\tfrac{1}{4}\sigma_1 w_{y})w_{t}-\tfrac{1}{24}\kappa \sigma_1 w_{x}^5-\tfrac{1}{3}w_{x}^3w_{y}
\\& \qquad
+(\tfrac{1}{4}\kappa w_{xx}^2-\tfrac{1}{4}\sigma_1 \kappa w_{y}^2)w_{x}
+\sigma_1 w_{xy}w_{xx} 
\big)
;
\end{aligned}&&
\end{flalign}
\end{subequations}
where $f_3(t)$, $f_4(t)$ are arbitrary functions. 
\end{thm}

\subsection{Conserved quantities}
Each of the conservation laws in Theorem~\ref{conslaws-mkp}
yields a conserved (time-independent) topological charge
\begin{equation}
\int_{\partial\Omega} (X,Y)\cdot\hat\nvec\,ds =0
\end{equation}
where the arbitrary function of $t$ appearing in $(X,Y)$
can be omitted without loss of generality. 
These charges can be used to introduce corresponding spatial potential systems:
\begin{equation}
X=\phi_y,
\quad
Y=-\phi_x,
\end{equation}
where $\phi$ is a potential. 

From conservation laws \eqref{mkp-conslaw1} and \eqref{mkp-conslaw2}, 
we obtain, respectively, 
\begin{equation}\label{mkp-charg1}
\begin{aligned}
\phi_y = &
w_{xxx} +\kappa w_x w_y +\tfrac{1}{3}\sigma_1 w_x^3 +w_t 
, 
\\
\phi_x = &
-\sigma_2 w_y +\tfrac{1}{2}\kappa w_x^2
,
\end{aligned}
\end{equation}
and 
\begin{equation}\label{mkp-charg2}
\begin{aligned}
\phi_y = &
(-w_{txxx}-(\sigma_1 w_{x}^2+\kappa w_{y})w_{tx}-\kappa w_{x}w_{ty}-w_{tt})y
+\kappa w_{x}w_{xxx} -\tfrac{1}{2}\kappa w_{xx}^2
\\& \qquad
+\tfrac{1}{4}\kappa w_{x}^4\sigma_1 +\tfrac{1}{2}\kappa^2w_{x}^2w_{y}-\tfrac{1}{2}\kappa \sigma_2 w_{y}^2
,
\\
\phi_x = &
(-\kappa w_{x}w_{tx}+\sigma_2 w_{ty})y +\tfrac{1}{6}\kappa ^2w_{x}^3 -\kappa \sigma_2 w_{x}w_{y} -\sigma_2 w_{t}
.
\end{aligned}  
\end{equation}
These two spatial potential systems hold for arbitrary $\kappa$. 

In the case of the mKP equation \eqref{mKP-case}, 
from conservation law \eqref{mkp-conslaw3}, we have 
\begin{equation}\label{mkp-charg3}
\begin{aligned}
\phi_y = &
(w_{xxx}+\tfrac{1}{3}\sigma_1 w_{x}^3+\kappa w_{y}w_{x}+w_{t})x
-\tfrac{1}{2}(\sigma_1 w_{txxx}
+(w_{x}^2+\kappa \sigma_1 w_{y})w_{tx} +\sigma_1 w_{tt}
\\& \qquad
+\sigma_1 \kappa w_{x}w_{ty})y^2
+(\sigma_1 w_{x}^2w_{y}+\sigma_1 \kappa w_{x}w_{xxx}-\tfrac{1}{2}\sigma_1 \kappa w_{xx}^2+\tfrac{1}{4}\kappa w_{x}^4+\tfrac{1}{2}\kappa w_{y}^2)y
-w_{xx}
,
\\
\phi_x = &
(\sigma_1 w_{y} + \tfrac{1}{2}\kappa w_{x}^2)x
-(\tfrac{1}{2}\sigma_1 \kappa w_{x}w_{tx}+\tfrac{1}{2}w_{ty})y^2
+(w_{t} +\kappa w_{y}w_{x} +\tfrac{1}{3}\sigma_1 w_{x}^3)y
,
\end{aligned}    
\end{equation}
while from conservation law \eqref{mkp-conslaw4}, we have 
\begin{equation}\label{mkp-charg4}
\begin{aligned}
\phi_y = &
(-\tfrac{1}{8}\kappa w_{ttt} -\tfrac{1}{8}w_{ttxxx}\kappa
-(\tfrac{1}{4}w_{y}+\tfrac{1}{8}\sigma_1 w_{x}^2\kappa )w_{ttx}
-\tfrac{1}{4}w_{tty}w_{x} -\tfrac{1}{2}w_{tx}w_{ty}
\\& \qquad
-\tfrac{1}{4} \sigma_1 \kappa w_{x}w_{tx}^2 )xy
+( \tfrac{1}{4}w_{x}w_{txxx}-\tfrac{1}{4}w_{txx}w_{xx}
+\tfrac{1}{4}(\kappa w_{y}w_{x}+w_{xxx}+\sigma_1 w_{x}^3)w_{tx}
\\& \qquad
+\tfrac{1}{8}\kappa w_{x}^2w_{ty}+\tfrac{1}{4}\sigma_1 w_{y}w_{ty} )x
+(\tfrac{1}{48}\kappa \sigma_1 w_{tttt}
+\tfrac{1}{48}\kappa\sigma_1 w_{tttxxx}
+(\tfrac{1}{24}\sigma_1 w_{y}+\tfrac{1}{48}\kappa w_{x}^2)w_{tttx}
\\& \qquad
+\tfrac{1}{24}\sigma_1 w_{x}w_{ttty}
+\tfrac{1}{8}(\kappa w_{x}w_{tx}+\sigma_1 w_{ty})w_{ttx}
+\tfrac{1}{8}\sigma_1 w_{tx}w_{tty} +\tfrac{1}{24}\kappa w_{tx}^3 )y^3
\\& \qquad
+( -\tfrac{1}{8}\sigma_1 (w_{ttxxx}w_{x} -w_{ttxx}w_{xx})
-\tfrac{1}{8}(w_{x}^3 +\sigma_1 w_{xxx} +\sigma_1 \kappa w_{y}w_{x})w_{ttx}
-\tfrac{1}{16}\sigma_1\kappa w_{tty}w_{x}^2
\\& \qquad
-\tfrac{1}{8}w_{tty}w_{y}
-\tfrac{1}{4}\sigma_1 w_{txxx}w_{tx}
+\tfrac{1}{8}\sigma_1 w_{txx}^2
-(\tfrac{1}{8}\kappa \sigma_1 w_{y}+\tfrac{3}{8}w_{x}^2)w_{tx}^2
-\tfrac{1}{4}\sigma_1 \kappa w_{tx}w_{ty}w_{x}
\\& \qquad
-\tfrac{1}{8}w_{ty}^2 )y^2
+(\tfrac{1}{8}\kappa w_{ttxx}) y
-\tfrac{1}{4}w_{txx}w_{x}+\tfrac{1}{2}w_{xx}w_{tx}+\tfrac{1}{8}w_{t}^2
+(\tfrac{1}{4}w_{xxx}+\tfrac{1}{12}\sigma_1 w_{x}^3
\\& \qquad
+\tfrac{1}{4}\kappa w_{y}w_{x})w_{t}
+\tfrac{1}{2}w_{xxx}^2
+\tfrac{1}{3}w_{xxx}\sigma_1 w_{x}^3
+(\tfrac{1}{2}\kappa w_{xxx}w_{y}-\tfrac{1}{2}\kappa w_{xx}w_{xy})w_{x}
-\tfrac{1}{2}\sigma_1 w_{xy}^2
\\& \qquad
+\tfrac{1}{4}\kappa w_{xx}^2w_{y} -\sigma_1 w_{xx}w_{yy}
+\tfrac{1}{18}w_{x}^6+\tfrac{5}{24}\kappa \sigma_1 w_{y}w_{x}^4 +\tfrac{1}{2}w_{y}^2w_{x}^2 +\tfrac{1}{12}\kappa \sigma_1w_{y}^3 
,
\\
\phi_x = &
-(\tfrac{1}{4}w_{ttx}w_{x}+\tfrac{1}{8}\sigma_1 \kappa w_{tty} +\tfrac{1}{4}w_{tx}^2 )xy
+( (\tfrac{1}{8}\kappa w_{x}^2 +\tfrac{1}{4}\sigma_1 w_{y})w_{tx}
+\tfrac{1}{8}\kappa \sigma_1 w_{tt} +\tfrac{1}{4}\sigma_1 w_{x}w_{ty} )x
\\& \qquad
+(\tfrac{1}{48}\kappa w_{ttty} +\tfrac{1}{24}\sigma_1 w_{tttx}w_{x} +\tfrac{1}{8}\sigma_1 w_{ttx}w_{tx} )y^3
-( \tfrac{1}{16}\kappa w_{ttt}
+(\tfrac{1}{16}\sigma_1 w_{x}^2\kappa +\tfrac{1}{8}w_{y})w_{ttx}
\\& \qquad
+\tfrac{1}{8}w_{tty}w_{x} 
+\tfrac{1}{8}\sigma_1 \kappa w_{tx}^2 w_{x} +\tfrac{1}{4}w_{tx}w_{ty} )y^2
+(\tfrac{1}{8}\kappa w_{x}^2 +\tfrac{1}{4}\sigma_1 w_{y})w_{t}
+\tfrac{1}{24}\kappa \sigma_1 w_{x}^5 +\tfrac{1}{3}w_{x}^3w_{y}
\\& \qquad
-(\tfrac{1}{4}\kappa w_{xx}^2-\tfrac{1}{4}\sigma_1 \kappa w_{y}^2)w_{x}
-\sigma_1 w_{xy}w_{xx}
.
\end{aligned}      
\end{equation}

It is surprising that,
in contrast to the situation for the KP equation considered in Ref.~\cite{AncGanRec2018}, 
there are no dynamical (non-topological) conserved quantities
admitted by the mKP family in potential form \eqref{mKP-pot-scaled}.

\section{Line soliton solutions}\label{solns}

A line soliton is a solitary wave in two dimensions,
\begin{equation}\label{linesoliton}
u=U(x+\mu y-\nu t) 
\end{equation}
with
\begin{equation}\label{linesoliton-bc}
U,U',U'',\text{etc.} \to 0 \text{ as } |x|,|y|\to\infty, 
\end{equation}  
where the parameters $\mu$ and $\nu$ determine the direction and the speed of the wave.

A more geometrical form for a line soliton is given by writing 
$x+\mu y= (x,y)\cdot\kvec$ 
with $\kvec=(1,\mu)$ being a constant vector in the $(x,y)$-plane. 
The travelling wave variable can then be expressed as 
\begin{equation}
\xi = x+\mu y-\nu t = |\kvec|( \hat\kvec\cdot (x,y) - c t )
\end{equation}
where the unit vector 
\begin{equation}
\hat\kvec = (\cos\theta,\sin\theta),
\quad
\tan\theta = \mu
\end{equation}
gives the direction of propagation of the line soliton,
and the constant 
\begin{equation}
c = \nu/|\kvec|, 
\quad
|\kvec|^2 = 1+\mu^2
\end{equation}
gives the speed of the line soliton. 
Since the direction of propagation stays the same under 
changing the direction angle by $\pm\pi$ while simultaneously changing the sign of the speed,
we will take the domain of $\theta$ to be $-\tfrac{1}{2}\pi < \theta \leq \tfrac{1}{2}\pi$. 

We will now derive the explicit line soliton solutions \eqref{linesoliton}
for the scaled mKP family \eqref{mKP-fam-scaled}. 
It will be convenient to use the coordinate form of the travelling wave variable
$\xi = x+\mu y-\nu t$ for this derivation.
Thus, we have $u_x = U'$, $u_y=\mu U'$, $u_t = -\nu U'$, and so on,
while $\partial_x^{-1}u_y = \mu\partial_\xi^{-1} U' = \mu U$ 
by the solitary wave conditions \eqref{linesoliton-bc}. 
Substitution of the line soliton expression \eqref{linesoliton} into equation \eqref{mKP-fam-scaled}
yields a nonlinear fourth-order ODE
\begin{equation}\label{mKP-U-linesoliton-ode}
(\sigma_2\mu^2 -\nu) U''+ \sigma_1 (U^2 U')' +\kappa\mu (U U')' + U''''=0 . 
\end{equation}
We can straightforwardly integrate this ODE twice to obtain a second-order ODE, 
and then we can use an integrating factor $U'$ to obtain a separable first-order ODE
\begin{equation}\label{mKP-U-energy}
U'{}^2  = \big( (\nu-\sigma_2\mu^2) - \tfrac{1}{3}\kappa\mu U -\tfrac{1}{6}\sigma_1 U^2 \big) U^2
\end{equation}
after use of conditions \eqref{linesoliton-bc}.

\begin{prop}
The general line soliton solution of the scaled mKP family \eqref{mKP-fam-scaled} 
is given by 
\begin{subequations}\label{mkp-fam-solitary}
\begin{equation}\label{mkp-fam-u}
u=\dfrac{6(\nu -\sigma_2\mu^2)}
{\sqrt{6\sigma_1\nu +(\kappa^2-6\sigma_1\sigma_2)\mu^2}\cosh(\sqrt{\nu -\sigma_2\mu^2}(x+\mu y-\nu t)) +\kappa\mu}
\end{equation}
where
\begin{equation}\label{mkp-fam-cond}
\nu-\sigma_2\mu^2 >0 \text{ if } \sigma_1=1; 
\quad
0<\nu -\sigma_2\mu^2 < \tfrac{1}{6}\kappa^2\mu^2
\text{ and } 
\mu >0 \text{ if } \sigma_1=-1 . 
\end{equation}
\end{subequations}
\end{prop}

With respect to the $x$ axis, 
the angle $\theta$ of the direction of motion of the line soliton
is given by $\arctan(\mu)$, 
while the speed of the line soliton is given by $\nu/\sqrt{1+\mu^2}$. 
These two parameters obey the kinematic condition \eqref{mkp-fam-cond}
which depends crucially on the signs of $\sigma_1$ and $\sigma_2$. 

In the case \eqref{mKP-case} representing the scaled mKP equation \eqref{mKP-scaled},
the general line soliton solution \eqref{mkp-fam-solitary} becomes 
\begin{subequations}\label{mkp-solitary}
\begin{equation}\label{mkp-u}
u=\dfrac{3\sqrt{2}(\nu +\sigma_1\mu^2)}
{ \sqrt{3\sigma_1\nu +4\mu^2}\cosh(\sqrt{\nu +\sigma_1\mu^2}(x+\mu y-\nu t)) +\mu }
\end{equation}
with the kinematic condition 
\begin{equation}\label{mkp-cond}
\nu>-\mu^2 \text{ if } \sigma_1=-\sigma_2=1; 
\quad
\mu^2<\nu <\tfrac{4}{3}\mu^2
\text{ and } 
\mu >0 \text{ if } \sigma_1=-\sigma_2=-1 . 
\end{equation}
\end{subequations}

We will next discuss a few properties of the mKP family of line solitons \eqref{mkp-fam-solitary}
in comparison to the mKP line solitons \eqref{mkp-solitary}.

\subsection{Subfamily containing the mKP equation} 

\begin{figure}[h]
\centering
\begin{subfigure}[t]{.4\textwidth}
\includegraphics[width=\textwidth]{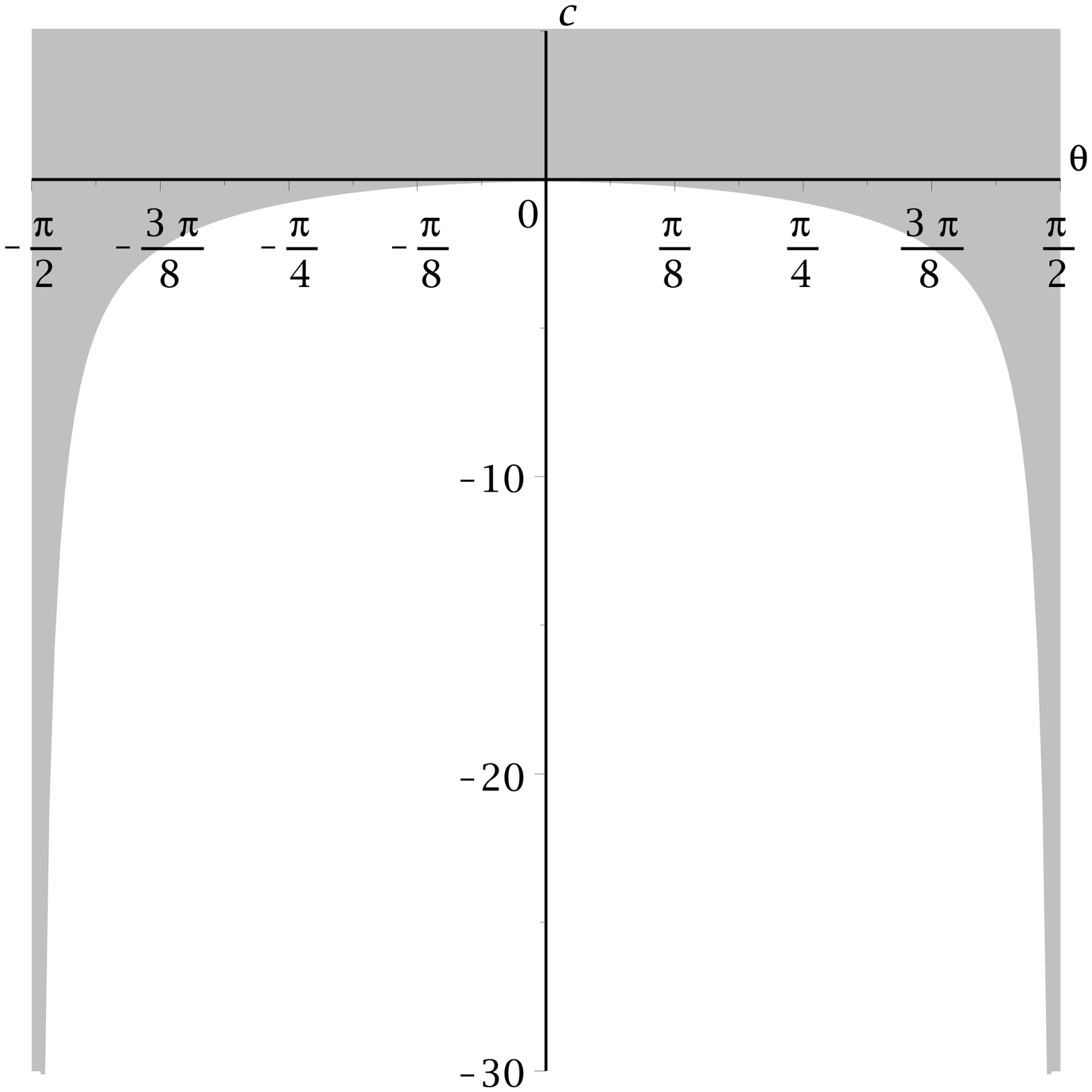}
\captionof{figure}{$\sigma_1=1$ (focussing)}
\label{fig:mkp-foc-kinregion}
\end{subfigure}%
\begin{subfigure}[t]{.4\textwidth}
\includegraphics[width=\textwidth]{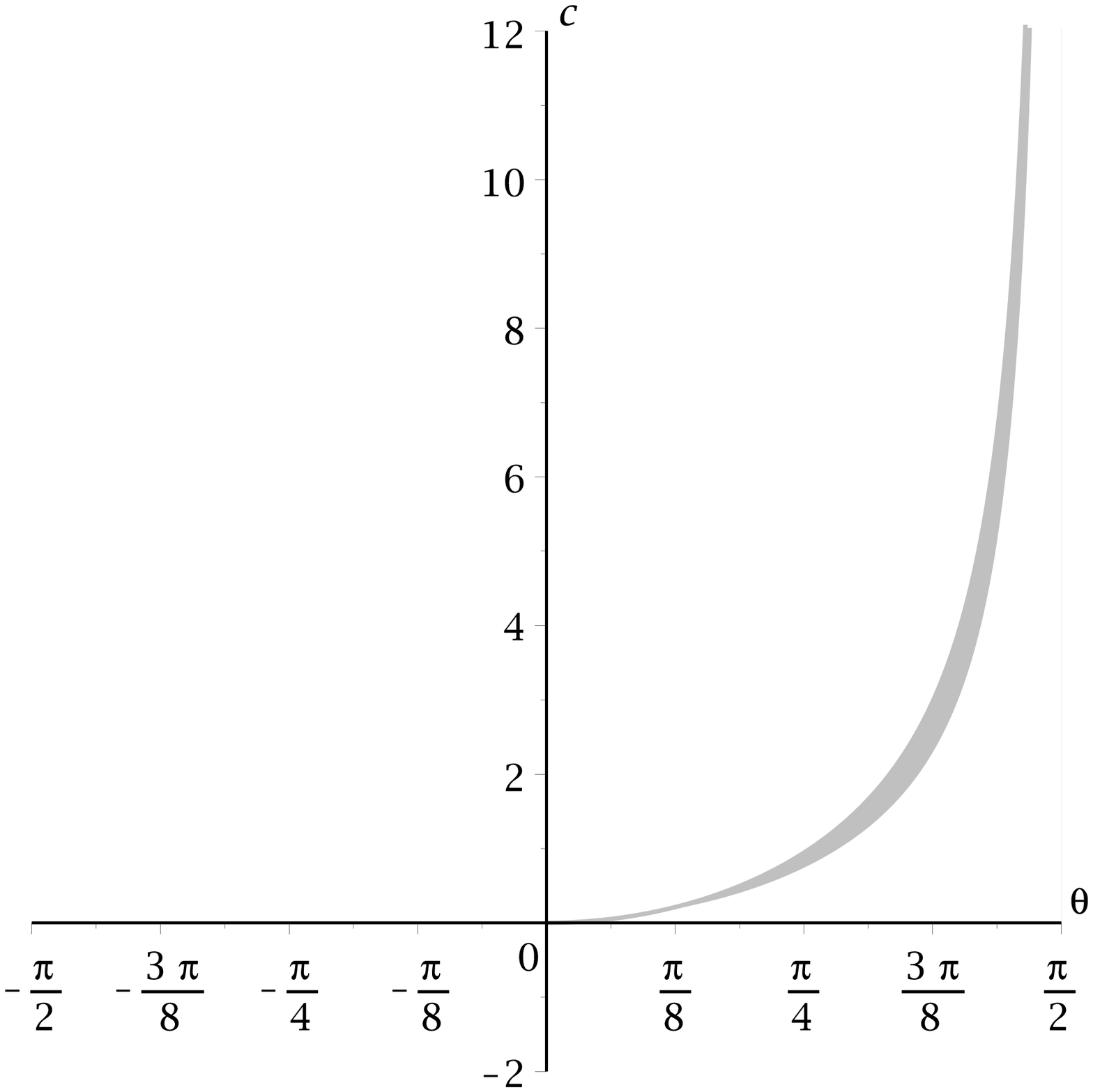}
\captionof{figure}{$\sigma_1=-1$ (defocussing)}
\label{fig:mkp-defoc-kinregion}
\end{subfigure}%
\caption{Kinematically allowed region in $(c,\theta)$ for the mKP family line soliton \eqref{mkpfam-incl-solitary}}
\end{figure}

\begin{figure}[h]
\centering
\begin{subfigure}[t]{.4\textwidth}
\includegraphics[width=\textwidth]{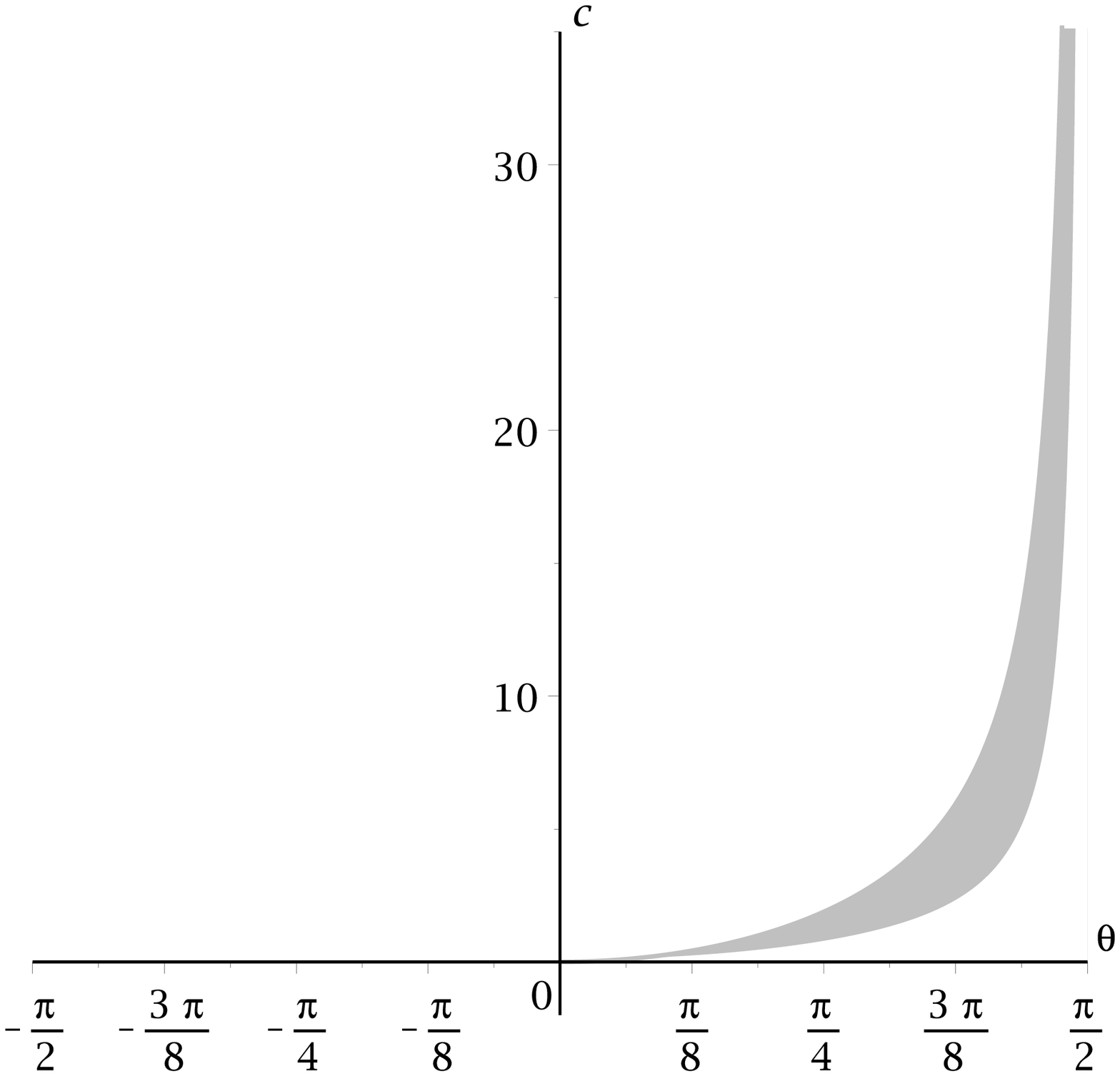}
\captionof{figure}{$\kappa^2=10$}
\label{fig:mkp-fam-kinregionfig-sigma1isneg1-sigma2is1-ksqis10}
\end{subfigure}%
\begin{subfigure}[t]{.4\textwidth}
\includegraphics[width=\textwidth]{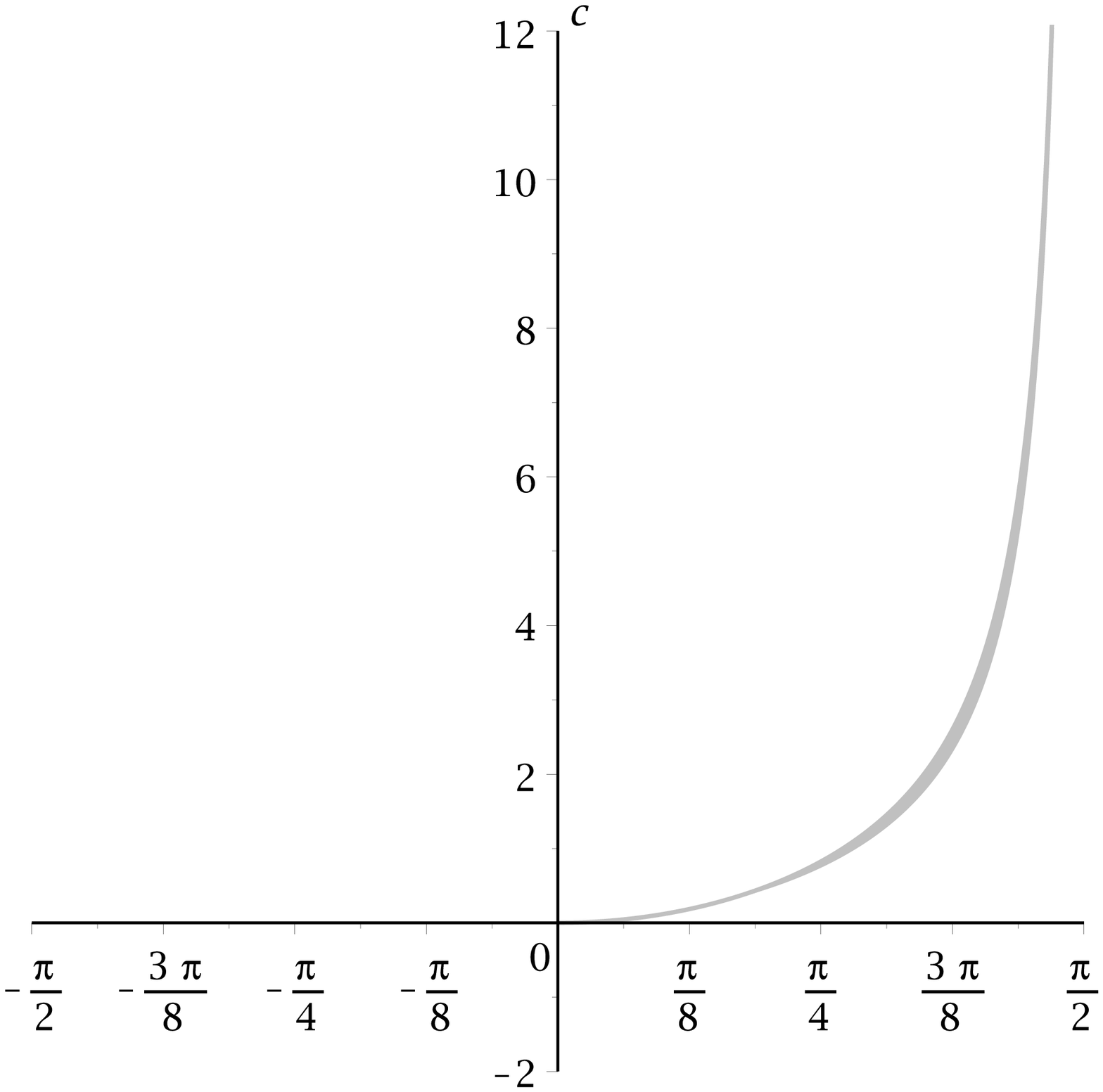}
\captionof{figure}{$\kappa^2=1$}
\label{fig:mkp-kinregionfig-sigma1is1-sigma2is1-ksqis1}
\end{subfigure}
\caption{Kinematically allowed region in $(c,\theta)$ for the mKP family line soliton \eqref{mkpfam-incl-solitary} in the defocussing case ($\sigma_1=-1$)}
\label{fig:mkpfam-incl-defoc-kinregion}
\end{figure}

To begin, we examine the case $\sigma_1\sigma_2=-1$,
where the mKP family constitutes a one-parameter ($\kappa$) extension of the mKP equation.
The line soliton \eqref{mkp-fam-solitary} in this case is given by 
\begin{equation}\label{mkpfam-incl-solitary}
u=\dfrac{6(\nu +\sigma_1\mu^2)}
{ \sqrt{6\sigma_1\nu +(\kappa^2 +6)\mu^2}\cosh(\sqrt{\nu +\sigma_1\mu^2}(x+\mu y-\nu t)) +\kappa\mu },
\quad
\sigma_1=\pm1,
\quad
\kappa>0 ,
\end{equation}
with the kinematic conditions
\begin{align}\label{mkpfam-incl-conds}
& \nu>-\mu^2 \text{ if } \sigma_1=1 , 
\\
& \mu^2 < \nu < (\tfrac{1}{6}\kappa^2+1)\mu^2
\text{ and } 
\mu >0 \text{ if } \sigma_1=-1 . 
\end{align}
In the focussing case, $\sigma_1=1$, 
there is a minimum negative speed $c>-\mu^2/\sqrt{1+\mu^2}$
which is a function of the angle $\theta=\arctan(\mu)$,
while there is no maximum speed. 
These minimum and maximum speeds are independent of $\kappa$. 
In the defocussing case, $\sigma_1=-1$, 
the speed has both a positive minimum and maximum, 
$\mu^2/\sqrt{1+\mu^2}<c<(1+\tfrac{1}{6}\kappa^2)\mu^2/\sqrt{1+\mu^2}$,
which depends on $\kappa$
where $\kappa^2=2$ recovers the mKP equation. 
The kinematically allowed region in $(c,\theta)$ is plotted 
in Fig.~\ref{fig:mkp-foc-kinregion} for the focussing case 
and in Figs.~\ref{fig:mkp-defoc-kinregion} and~\ref{fig:mkpfam-incl-defoc-kinregion}
for the defocussing case. 

\begin{figure}[h]
\centering
\begin{subfigure}[t]{.4\textwidth}
\includegraphics[width=\textwidth]{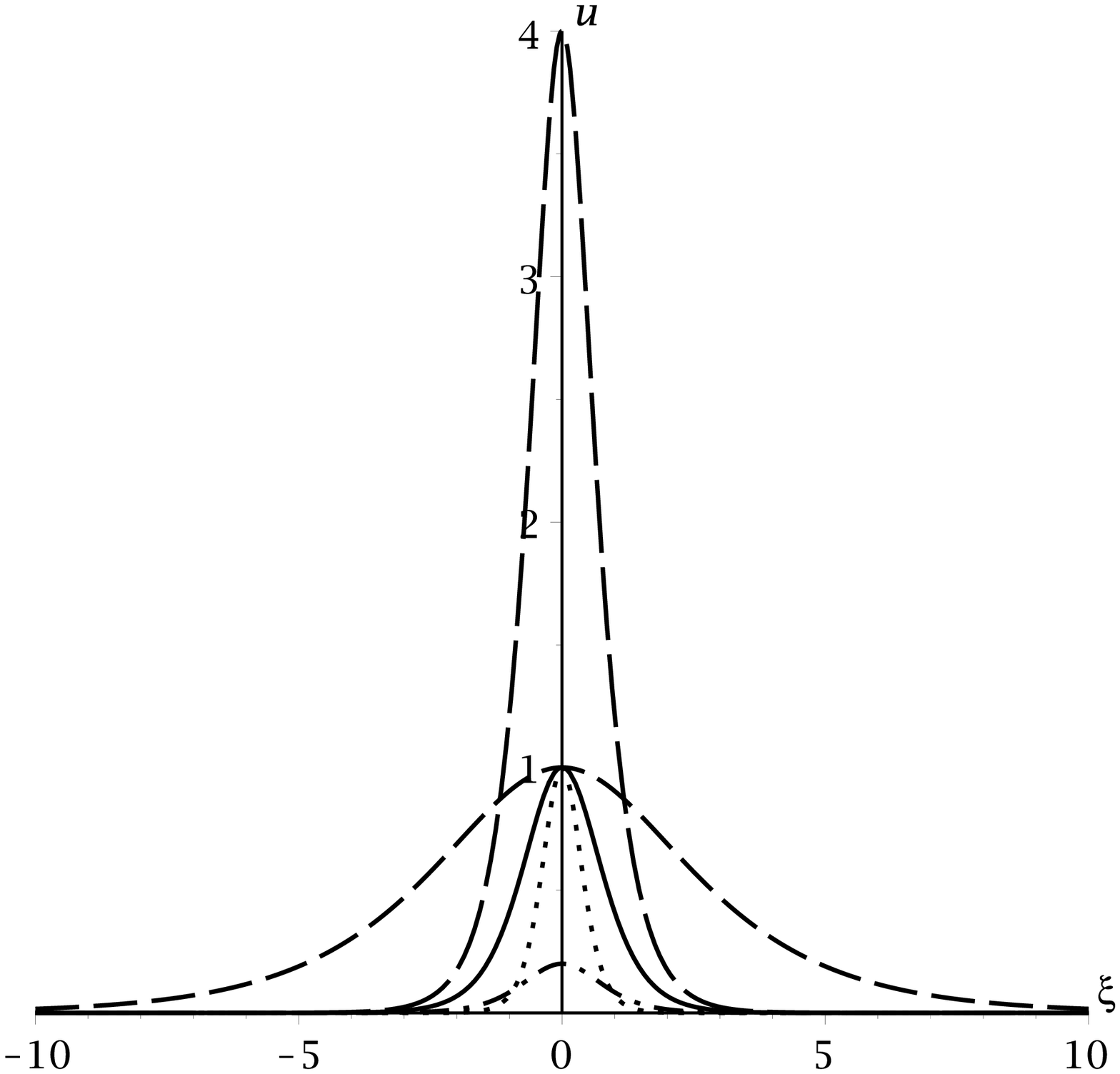}
\captionof{figure}{$\sigma_1=1$ (focussing)}
\label{fig:mkp-profilefig-sigma1is1}
\end{subfigure}%
\begin{subfigure}[t]{.4\textwidth}
\includegraphics[width=\textwidth]{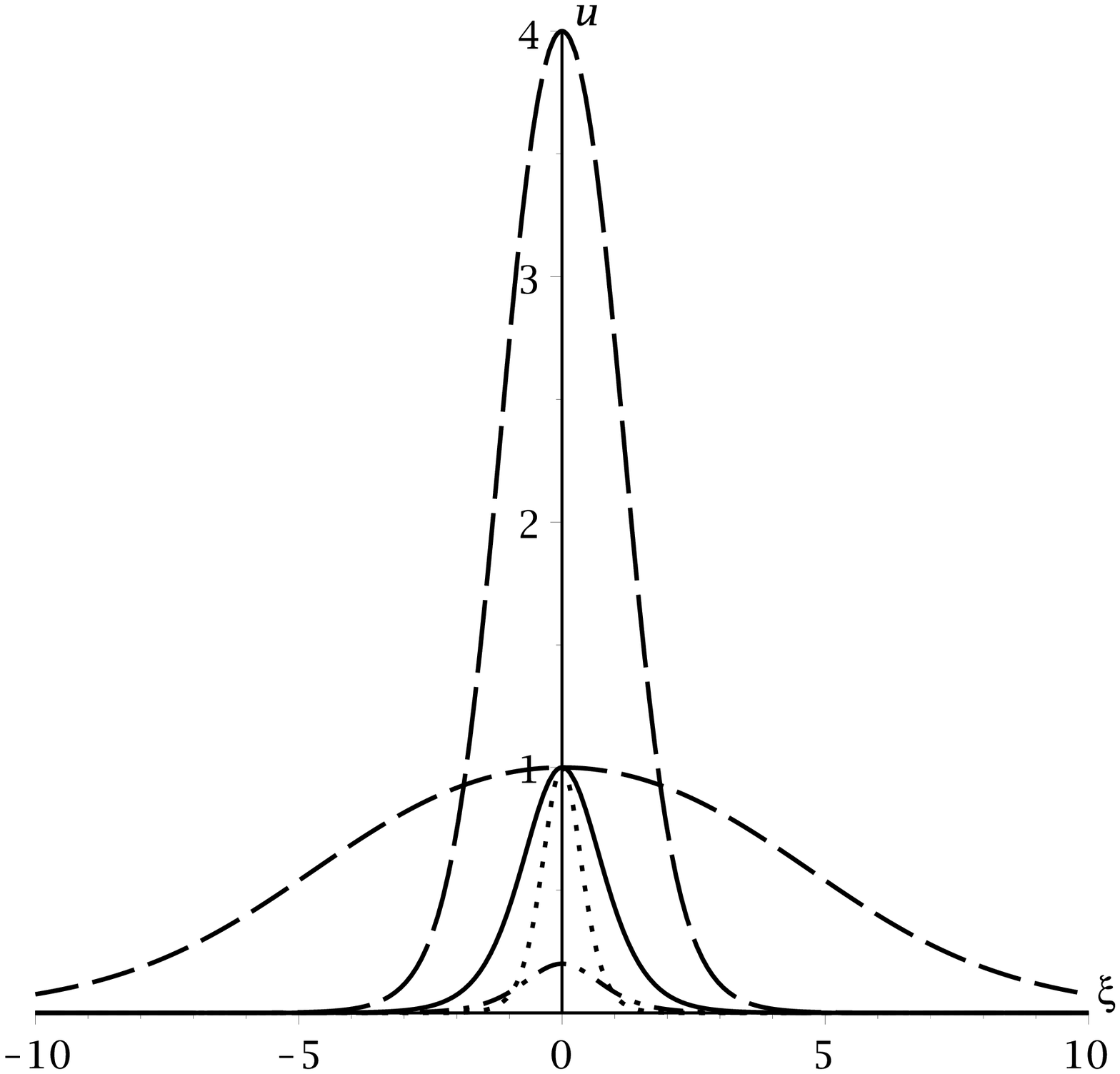}
\captionof{figure}{$\sigma_1=-1$ (defocussing)}
\label{fig:mkp-profilefig-sigma1isneg1}
\end{subfigure}
\caption{Profile of the mKP line soliton \eqref{mkp-solitary} 
and mKP family line soliton \eqref{mkpfam-incl-solitary} and \eqref{mkpfam-excl-solitary}
for $(h,w)=$ 
$(1,1)$ (solid); $(4,1)$ (long dash); $(\tfrac{1}{5},1)$ (dash dot); $(1,4)$ (dash); $(1,\tfrac{1}{2})$ (dot)}
\label{fig:mkp-profilefig}
\end{figure}

In the focussing case, 
the line soliton has height 
$h=6(\nu+\mu^2)/(\sqrt{(\kappa^2+6)\mu^2+6\nu}+\kappa\mu)$
and width (proportional to) $w=2/\sqrt{\nu+\mu^2}$. 
These expressions can be inverted and substituted into the line soliton \eqref{mkpfam-incl-solitary}, 
yielding 
\begin{equation}\label{mkpfam-foc-profile}
u= 24h/((h^2w^2+24)\cosh(\xi/w)^2 -h^2w^2)
\end{equation}
for the profile of the line soliton in terms of its height and width. 
Notice that it does not depend on $\kappa$ and hence it is the same as for the ordinary mKP line soliton. 
Plots of this profile are shown in Fig.~\ref{fig:mkp-profilefig-sigma1is1}. 

Likewise in the defocussing case, 
the height and width of the line soliton are 
$h=6(\nu-\mu^2)/(\sqrt{(\kappa^2+6)\mu^2-6\nu}+\kappa\mu)$
and $w=2/\sqrt{\nu-\mu^2}$. 
The profile of the line soliton in terms of its height and width is given by 
\begin{equation}\label{mkpfam-defoc-profile}
u= 24h/((24-h^2w^2)\cosh(\xi/w)^2 +h^2w^2),
\quad
hw< \sqrt{24} . 
\end{equation}
Plots of this profile are shown in Fig.~\ref{fig:mkp-profilefig-sigma1isneg1}. 
Notice again that it does not depend on $\kappa$ and hence it is the same as for the ordinary mKP line soliton. 
An interesting contrast compared to the focussing case is that the height and width 
must obey a kinematic condition. 

Overall, the kinematic properties of the line soliton \eqref{mkpfam-incl-solitary}
in non-integrable case $\kappa^2\neq 2$ are qualitatively the same as those in the integrable case $\kappa^2=2$.

\subsection{Subfamily excluding the mKP equation} 

\begin{figure}[h]
\centering
\begin{subfigure}[t]{.33\textwidth}
\includegraphics[width=\textwidth]{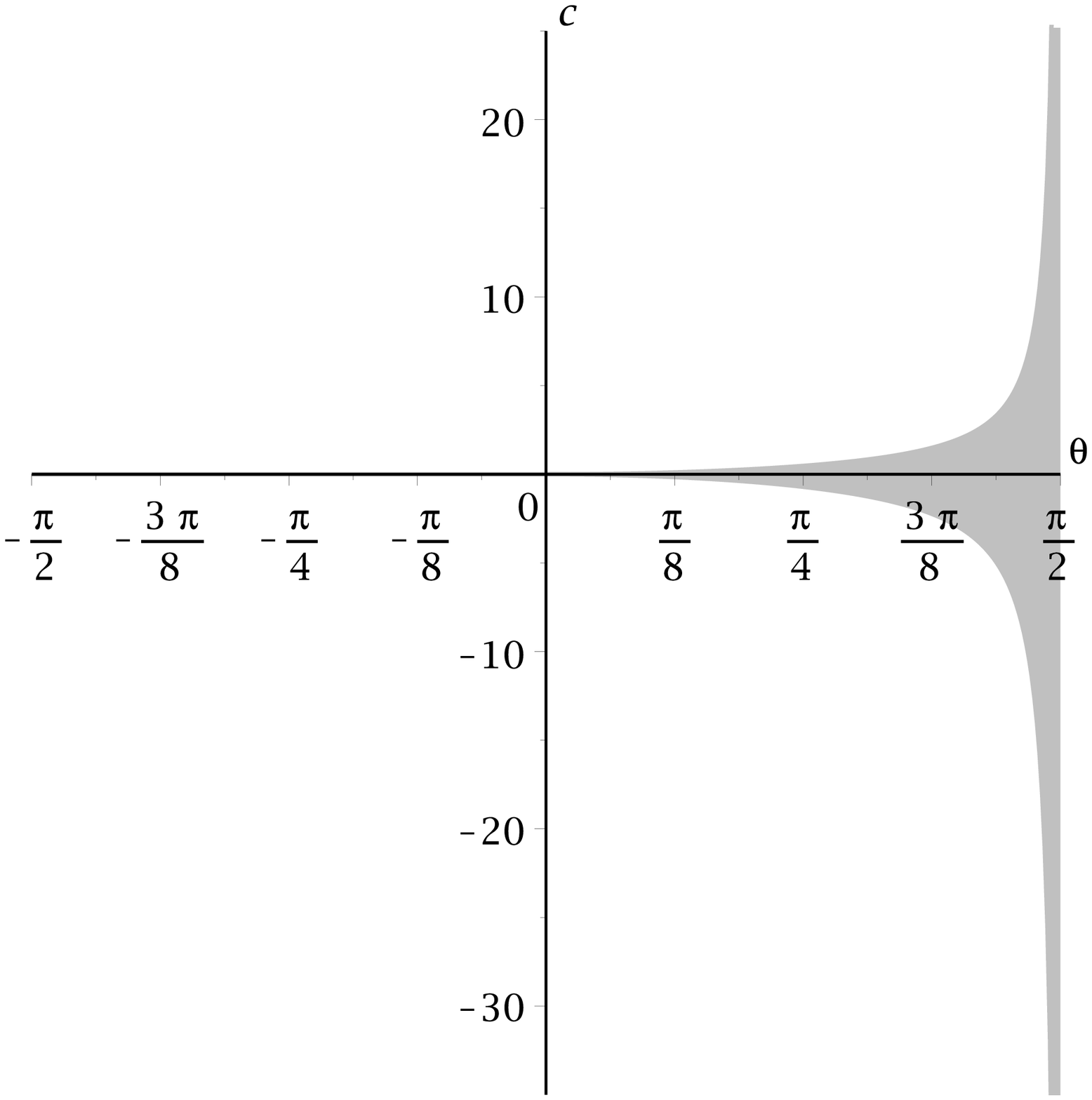}
\captionof{figure}{$\kappa^2=10$}
\label{fig:mkp-fam-kinregionfig-sigma1isneg1-sigma2isneg1-ksqis10}
\end{subfigure}%
\begin{subfigure}[t]{.33\textwidth}
\includegraphics[width=\textwidth]{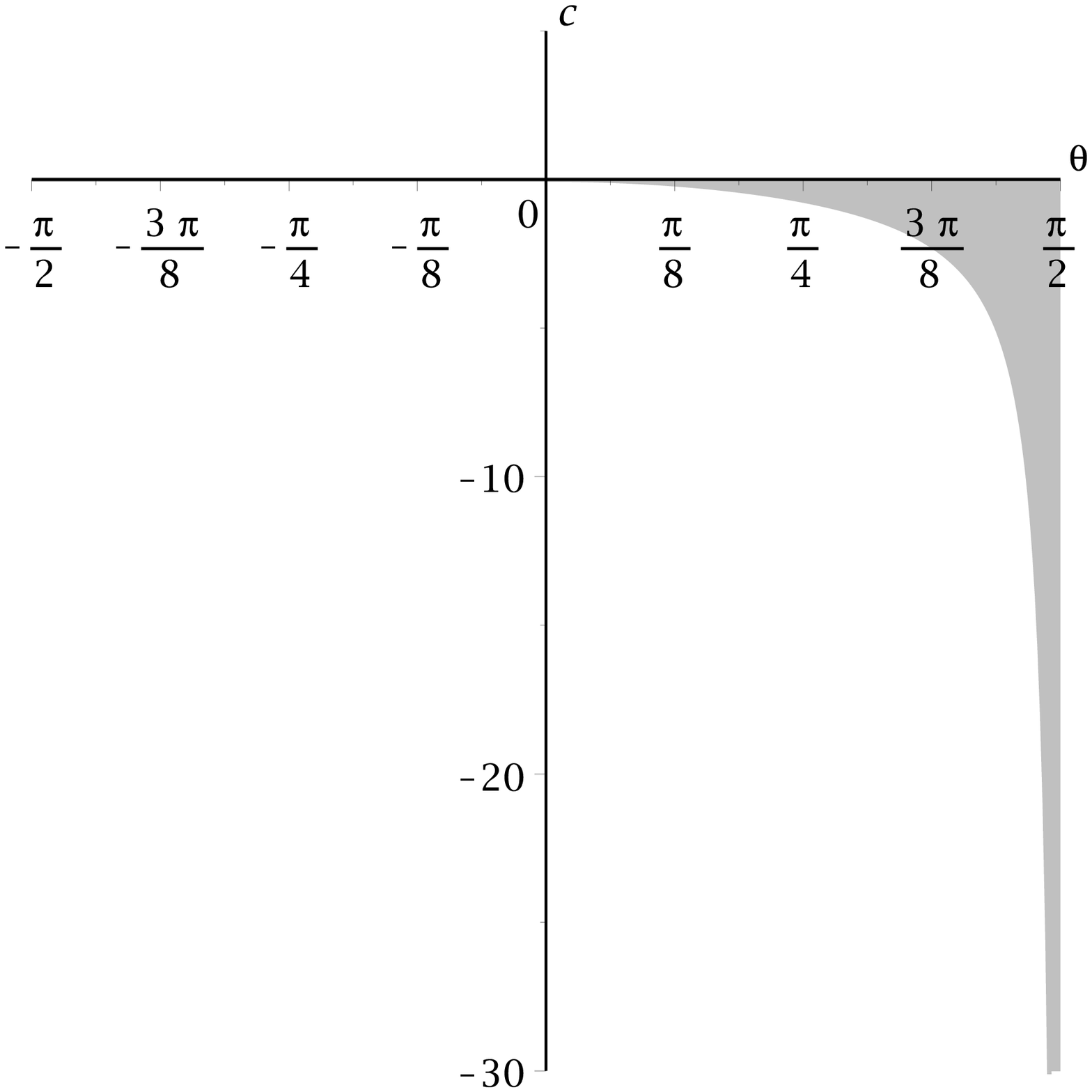}
\captionof{figure}{$\kappa^2=6$}
\label{fig:mkp-kinregionfig-sigma1is1-sigma2isneg1-ksqis6}
\end{subfigure}%
\begin{subfigure}[t]{.33\textwidth}
\includegraphics[width=\textwidth]{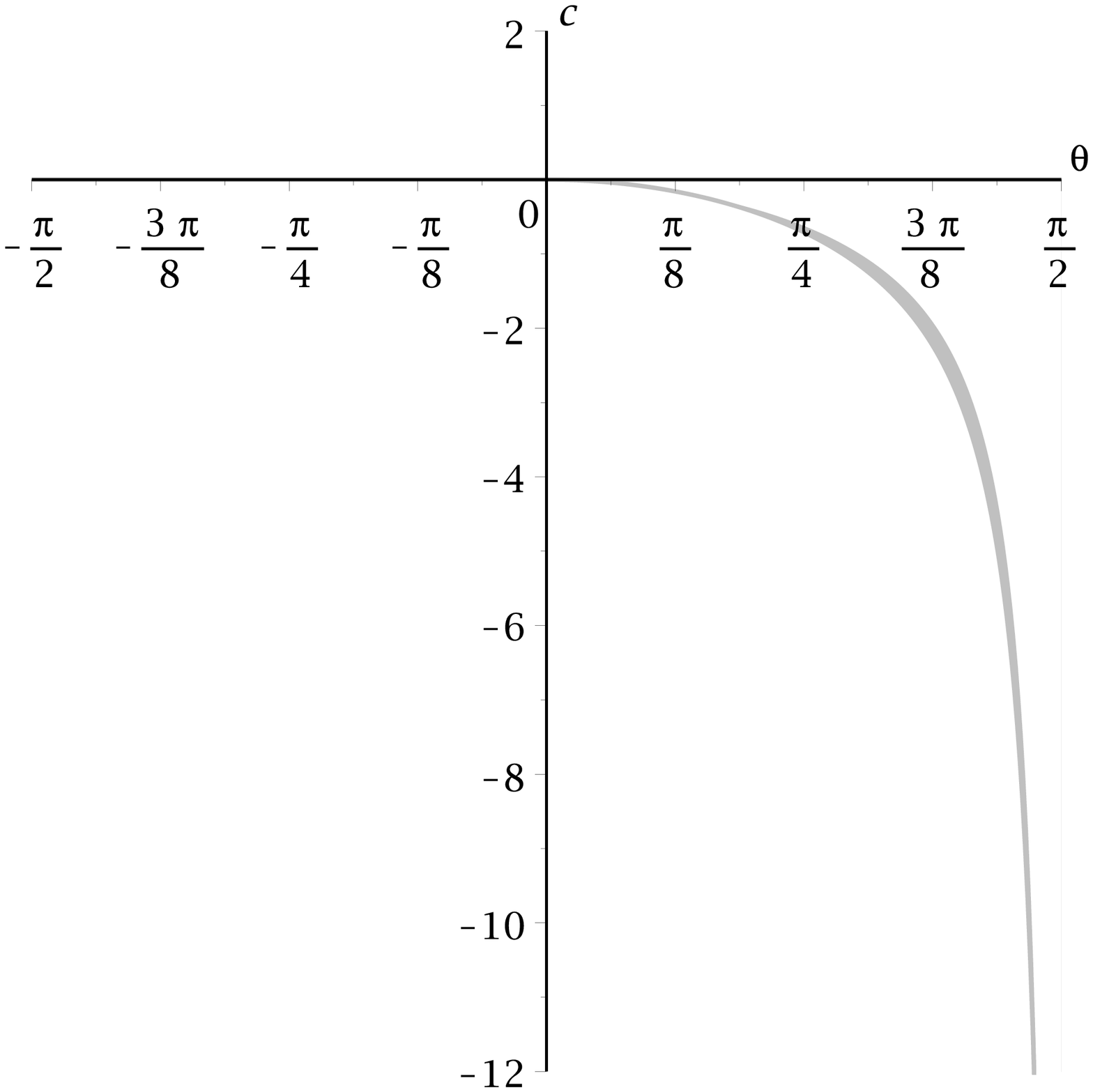}
\captionof{figure}{$\kappa^2=1$}
\label{fig:mkp-kinregionfig-sigma1is1-sigma2isneg1-ksqis1}
\end{subfigure}%
\caption{Kinematically allowed region in $(c,\theta)$ for the mKP family line soliton \eqref{mkpfam-excl-solitary} in the defocussing case ($\sigma_1=-1$)}
\label{fig:mkpfam-excl-defoc-kinregion}
\end{figure}

Last, we examine the case $\sigma_1\sigma_2=1$,
where the mKP family is a strict generalization of the mKP equation.
The line soliton \eqref{mkp-fam-solitary} in this case is given by 
\begin{equation}\label{mkpfam-excl-solitary}
u=\dfrac{6(\nu -\sigma_1\mu^2)}
{ \sqrt{6\sigma_1\nu +(\kappa^2 -6)\mu^2}\cosh(\sqrt{\nu +\sigma_1\mu^2}(x+\mu y-\nu t)) +\kappa\mu },
\quad
\sigma_1=\pm1,
\quad
\kappa>0 ,
\end{equation}
with the kinematic conditions
\begin{align}
& \nu>\mu^2 \text{ if } \sigma_1=1 , 
\\
& -\mu^2 < \nu < (\tfrac{1}{6}\kappa^2-1)\mu^2
\text{ and } 
\mu >0 \text{ if } \sigma_1=-1 . 
\end{align}
In the focussing case, $\sigma_1=1$, 
the speed has a positive minimum $c>\mu^2/\sqrt{1+\mu^2}$
and no maximum. 
In the defocussing case, $\sigma_1=-1$, 
there is a minimum negative speed $c>-\mu^2/\sqrt{1+\mu^2}$,
while the maximum speed $c<(\tfrac{1}{6}\kappa^2-1)\mu^2/\sqrt{1+\mu^2}$ 
is either positive if $\kappa^2>6$ or negative if $\kappa^2<6$.
Plots of the kinematically allowed region in $(c,\theta)$ 
are shown in Fig.~\ref{fig:mkp-foc-kinregion} for the focussing case 
and in Figs.~\ref{fig:mkp-defoc-kinregion} and~\ref{fig:mkpfam-excl-defoc-kinregion}
for the defocussing case. 
These kinematic properties of the line soliton \eqref{mkpfam-excl-solitary}
are qualitatively different compared to those in the (extended) mKP case \eqref{mkpfam-incl-solitary}. 

In the focussing case, 
the line soliton has height 
$h=6(\nu+\mu^2)/(\sqrt{(\kappa^2-6)\mu^2-6\nu}+\kappa\mu)$
and width (proportional to) $w=2/\sqrt{\nu+\mu^2}$. 
The profile of this line soliton in terms of its height and width
is the same as the profile \eqref{mkpfam-foc-profile} in the (extended) mKP case. 
Plots are shown in Fig.~\ref{fig:mkp-profilefig-sigma1is1}. 

Similarly in the defocussing case, 
the height and width of the line soliton are 
$h=6(\nu-\mu^2)/(\sqrt{(\kappa^2-6)\mu^2+6\nu}+\kappa\mu)$
and $w=2/\sqrt{\nu-\mu^2}$. 
The profile in terms of its height and width 
is the same as the profile \eqref{mkpfam-defoc-profile} in the (extended) mKP case. 
Plots are shown in Fig.~\ref{fig:mkp-profilefig-sigma1isneg1}.

\section{Concluding remarks}\label{remarks}

We have obtained in explicit form 
all of the line soliton solutions and all of the low-order conservation laws 
for the family \eqref{mKP-fam} of mKP equations. 

When $\alpha\gamma>0$,
the family includes the well-known integrable mKP equation \eqref{mKP}.
The line solitons in this case have qualitatively similar kinematic properties to the mKP line soliton.

In contrast, when $\alpha\gamma<0$,
the family is a strict generalization of the mKP equation \eqref{mKP}.
The kinematic properties of the line solitons in this case
are qualitatively different compared to the mKP line soliton.
In particular, in the focussing case $\alpha<0$,
the speed of the line soliton is strictly positive,
and in the defocussing case $\alpha>0$,
the speed will be strictly negative if $\kappa^2<6\alpha|\gamma|$. 

Our results can be used as a starting point
to investigate the stability of the line soliton solutions
and to determine whether their stability depends on integrability condition \eqref{mKP-constraint}.

\section{Acknowledgements}
S.C.A.\ is supported by an NSERC research grant
and thanks the University of C\'adiz for additional support during the period 
when this work was initiated.

\section*{Appendix}

The determining equation \eqref{Q-deteqn} for multipliers \eqref{low-order-Q} 
with differential order less than four 
splits with respect to the set of variables $\{\partial^4 w,\partial^5 w,\partial^6 w\}$. 
We have carried out the setting up and splitting of the determining equation 
by using Maple.
This yields an overdetermined system consisting of 3356 equations 
to be solved for $Q$ as well as for $\kappa\neq0$, 
with $\sigma_1^2=\sigma_2^2=1$. 
Solving the system is a nonlinear problem because 
$Q$ appears linearly in products with $\kappa$. 
We use the Maple package 'rifsimp' to find the complete case tree of solutions. 
For each solution case in the tree, we solve the system of equations 
by using Maple 'pdsolve' and 'dsolve', 
and we check that the solution has the correct number of free constants/functions
and satisfies the original overdetermined system. 
Finally, we merge overlapping cases by following the method explained in Ref.~\cite{RecAnc2017}.


\begin{thebibliography}{1}

\bibitem{AblSeg1979}
M.J. Ablowitz and H. Segur, 
On the evolution of packets of water waves, 
J. Fluid Mech. 92 (1979), 691--715. 


\bibitem{Anc-review}
S.C. Anco, 
Generalization of Noether's theorem in modern form to non-variational partial differential equations. 
In: Recent progress and Modern Challenges in Applied Mathematics, Modeling and Computational Science, 119--182, 
Fields Institute Communications, Volume 79 (2017). 

\bibitem{Anc2003}
S.C. Anco, 
Conservation laws of scaling-invariant field equations, 
J. Phys. A: Math. and Gen. 36 (2003), 8623--8638.

\bibitem{AncBlu2002b}
S. C. Anco and G. Bluman,
Direct construction method for conservation laws of partial differential equations Part II: General treatment,
Euro. J. Appl. Math. 41 (2002), 567--585.

\bibitem{AncGanRec2018}
S.C. Anco, M. Gandarias, E. Recio,
Conservation laws, symmetries, and line soliton solutions of generalized KP and Boussinesq equations with $p$-power nonlinearities in two dimensions, 
Theor. Math. Phys. 197(1) (2018), 1393--1411. 


\bibitem{BCA-book}
G.W. Bluman, A Cheviakov, S.C. Anco,
Applications of Symmetry Methods to Partial Differential Equations.
\emph{New York: Springer} (2009).   

\bibitem{CheLiu}
Y. Chen, P. L.-F. Liu, 
A generalized modified Kadomtsev-Petviashvili equation
for interfacial wave propagation near the critical depth level, 
Wave Motion 27 (1998), 321--339. 

\bibitem{DasSar}
G.C. Das, J. Sarma,
Evolution of solitary wave in multicomponent plasmas,
Chaos, Solitons and Fractals 9 (1998), 901--911. 

\bibitem{GesHolSaaSim}
F. Gesztesy, H. Holden, E. Saab, B. Simon, 
Explicit construction of solutions of the modified Kadomtsev-Petviashvili equation, 
J. Funct. Anal. 98(1) (1991), 211--228.

\bibitem{KadPet}
B.B. Kadomstev and V.I. Petviashvili,
On the stability of waves in weakly dispersive media, 
Sov. Phys. Dokl. 15 (1970), 539--541.

\bibitem{KonDub1984}
B. Konopelchenko, V. Dubrovsky, 
Some new integrable nonlinear evolution equations in 2+1 dimensions, 
Phys. Lett. A 102 (1984), 15--17. 

\bibitem{KonDub1992}
B.G. Konopelchenko and V.G. Dubrovsky,
Inverse spectral transform for the modified Kadomtsev-Petviashvili equation, 
Studies in Applied Math. 86(3) (1992), 219--268. 

\bibitem{NazAliNae}
R. Naz, Z. Ali, and I. Naeem, 
Reductions and New Exact Solutions of ZK, Gardner KP, and Modified KP Equations via Generalized Double Reduction Theorem. 
Abstract and Applied Analysis (2013), 340564--340575.

\bibitem{Olv-book}
P.J. Olver, 
{\em Applications of Lie Groups to Differential Equations},
Springer-Verlag, New York, 1993.

\bibitem{RecAnc2017}
E. Recio, S.C. Anco, 
Conservation laws and symmetries of radial generalized nonlinear $p$-Laplacian evolution equations, 
J. Math. Anal. Appl. 452 (2017) 1229--1261.

\bibitem{VeeDan}
V. Veerakumar and M. Daniel,
Modified Kadomtsev-Petviashvili (MKP) equation and electromagnetic soliton,
Math. Comput. Simulat. 62 (2003), 163--169.

\bibitem{Wol}
T. Wolf,
A comparison of four approaches to the calculation of conservation laws,
Euro. J. Appl. Math. 13 (2002), 129--152.

\bibitem{ZhaXuJiaZho}
X. Zhao, W. Xu, H. Jia, and H. Zhou, 
Solitary wave solutions for the modified Kadomtsev-Petviashvili equation, 
Chaos, Solitons and Fractals 34(2) (2007), 465--475. 





\end{thebibliography}
\end{document}